\documentclass{emulateapj}
\usepackage[]{natbib, graphicx}

\shorttitle{Do Moderate-Luminosity AGN Suppress Star Formation?}
\shortauthors{Schawinski et al.}
\slugcomment{Accepted for publication in the Astrophysical Journal Letters 12/24/08}
\begin{document}

\title{Do Moderate-Luminosity Active Galactic Nuclei Suppress Star Formation?}

\author{
Kevin Schawinski,\altaffilmark{1,2}
Shanil Virani,\altaffilmark{1,2}
Brooke Simmons,\altaffilmark{1,2}
C. Megan Urry,\altaffilmark{1,2}
Ezequiel Treister,\altaffilmark{3,4}
Sugata Kaviraj\altaffilmark{5,6}
and Bronika Kushkuley\altaffilmark{1,2}
}

\altaffiltext{1}{Department of Physics, Yale University, New Haven, CT 06511, U.S.A.}
\altaffiltext{2}{Yale Center for Astronomy and Astrophysics, Yale University, P.O. Box 208121, New Haven, CT 06520, U.S.A.}
\altaffiltext{3}{Chandra Fellow}
\altaffiltext{4}{Institute for Astronomy, 2680 Woodlawn  Drive, University of Hawaii, Honolulu, HI 96822, U.S.A.}
\altaffiltext{5}{Department of Physics, University of Oxford, Oxford OX1 3RH, UK.}
\altaffiltext{6}{Reseach Fellow funded by the Royal Commission for the 
Exhibition of 1851.}
\email{kevin.schawinski@yale.edu}

\begin{abstract}
The growth of supermassive black holes and their host galaxies are thought to be linked, but the precise nature of this symbiotic relationship is still poorly understood. Both observations and simulations of galaxy formation suggest that the energy input from active galactic nuclei (AGN), as the central supermassive black hole accretes material and grows, heats the interstellar material and suppresses star formation. In this Letter, we show that most host galaxies of moderate-luminosity supermassive black holes in the local universe have intermediate optical colors that imply the host galaxies are transitioning from star formation to quiescence, the first time this has been shown to be true for all AGN independent of obscuration. The intermediate colors suggest that star formation in the host galaxies ceased roughly 100 Myr ago. This result indicates that either the AGN are very long-lived, accreting for more than 1 Gyr beyond the end of star formation, or there is a $\sim$100 Myr time delay between the shutdown of star formation and detectable black hole growth. The first explanation is unlikely given current estimates for AGN lifetimes, so low-lumiosity AGN must shut down star formation before substantial black hole accretion activity is detected. The scarcity of AGN host galaxies in the blue cloud reported here challenges scenarios where significant star formation and black hole growth are coeval. Lastly, these observations also strongly support the `Unified Model' of AGN  as the host galaxy colors are independent of obscuration towards the central engine.
\end{abstract}


\section{Introduction}

Several works have noted that obscured AGN appear to be prevalent in the `green valley' on the color-magnitude diagram, between actively star-forming galaxies in the blue cloud and passively evolving galaxies on the red sequence \citep{2007MNRAS.382.1415S, 2007ApJS..173..267S, 2007ApJ...660L..11N, 2008ApJ...675.1025S, 2008ApJ...673..715C, 2008ApJ...681..931B, 2008MNRAS.385.2049G, 2009ApJ...690.1672S}. The presence of AGN in galaxies with such intermediate optical colors has been interpreted as evidence for the role of AGN in the suppression of star formation, so that AGN host galaxies are transitioning from the blue cloud to the red sequence. However, the obscuration that affects the central nuclear region might also be reddening the galaxy colors systematically. Therefore, to understand the connection between star formation and black hole growth, we need to study a complete, unbiased sample of AGN. 

In particular, do all AGN, including both unobscured and highly obscured, Compton-thick AGN, lie in the green valley, as one might hypothesize from the Unified Model \citep{1995PASP..107..803U, 1993ARA&A..31..473A}? In this \textit{Letter}, we test the universality of the association of AGN with `green' optical colors using a complete sample of obscured and unobscured AGN host galaxies at very low redshift and then discuss the implications of the observed optical color distribution for the r\^{o}le of moderate-luminosity AGN in the suppression of star formation. We include for the first time not just obscured AGN, but also unobscured and highly obscured sources, and can thus draw strong conclusions on the host galaxies of AGN - not just where on the color-magnitude relation they lie, but also where they are \textit{absent}.

\begin{figure*}
\begin{center}

\includegraphics[width=\textwidth]{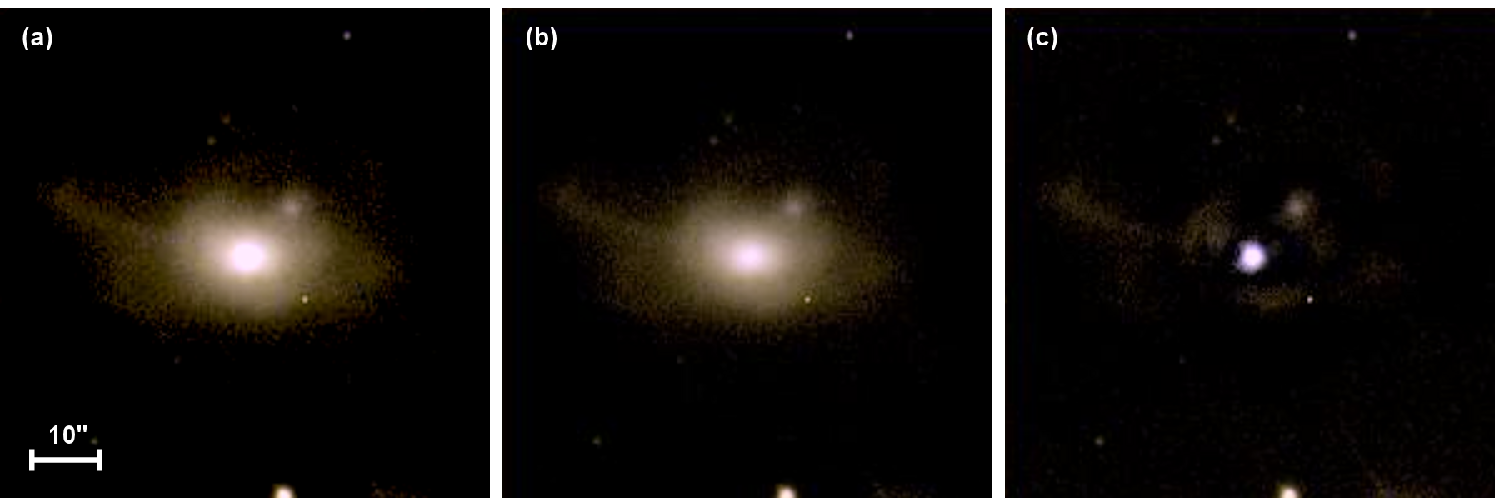}\\
\vspace{0.1in}
\includegraphics[width=\textwidth]{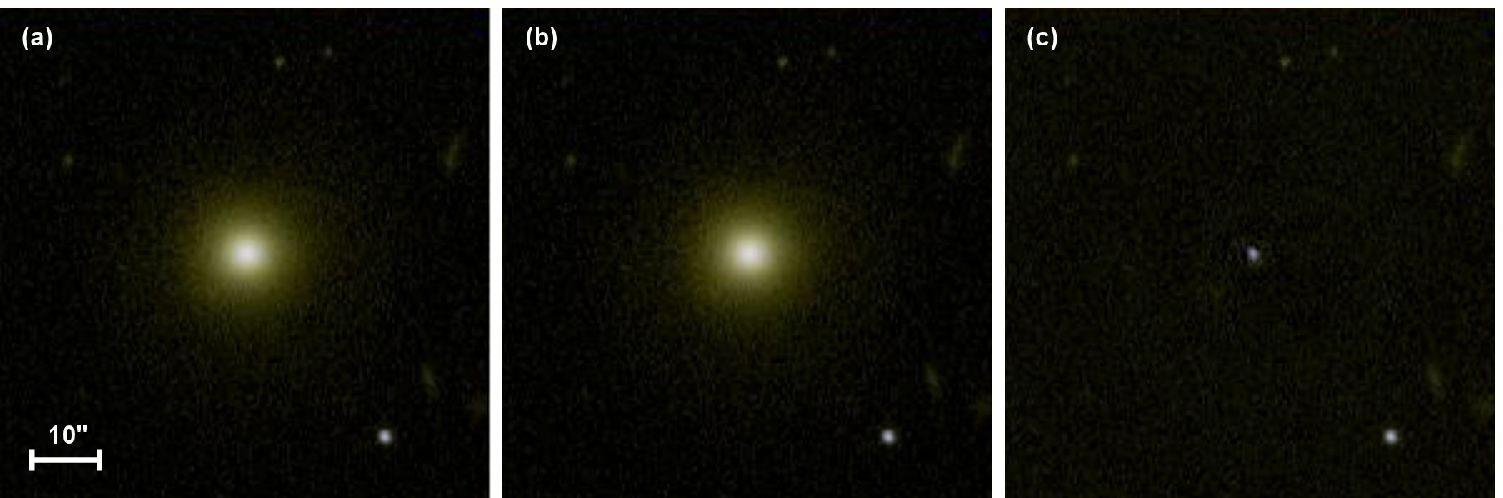}
\caption{Two examples of the central point source subtraction for SDSS $gr$ images of a morphologically disturbed galaxy \textit{(top row)} and an elliptical galaxy \textit{(bottom row)}. In panels \textit{(a)}, we show the original image. In panel \textit{(b)},  the central point source is subtracted. In panel \textit{(c)},  the galaxy model is subtracted. The residuals show the blue central point source, and in the case of the merger galaxy, tidal tails and shells.}

\label{fig:example}

\end{center}
\end{figure*}

\section{\textit{Swift} BAT Sample Selection}
In order to select a complete sample of both unobscured and obscured AGN, we use hard X-ray selection, which is nearly unbiased with respect to obscuration, yet remains highly efficient, unlike IR selection techniques. The Burst Alert Telescope (BAT) onboard the \textit{Swift} satellite \citep{2004ApJ...611.1005G} observes photons in the 14-195 keV range in which even the most highly obscured AGN can be detected. 

We start with the sample of AGN found by the \textit{Swift} BAT in the first 9 months of operation across the entire sky \citep{2008ApJ...681..113T}, then limit ourselves to objects that overlap with the Sloan Digital Sky Survey (SDSS; \citealt{2000AJ....120.1579Y, 2008ApJS..175..297A}) and have redshifts of $0.01 < z < 0.07$, yielding 21 AGN in the luminosity range of $L_{14-195~\rm keV} = 10^{42.7}-10^{44.5} ~\rm erg~s^{-1}$ with a median of $10^{44.0} ~\rm erg~s^{-1}$. The AGN include the entire range from entirely unobscured to highly obscured, and have moderate luminosities, and thus accretion rates, typical of the local universe. Quasars exhibit even higher luminosities and accretion rates, but are rare in the low redshift universe; our sample probes almost all of the significant black hole growth in the local universe.

\section{Host Galaxy Colors and Central Point Source Subtraction}
To study the host galaxy optical properties, we use imaging data from the SDSS DR6 \citep{2000AJ....120.1579Y, 2008ApJS..175..297A}. The main challenge in understanding the host galaxy colors of AGN is the contamination of the non-stellar central point source of the AGN. Deep X-ray surveys can deliver samples of X-ray selected AGN down to moderate X-ray luminosities out to considerable redshifts, but the contamination by the AGN light can make comparing AGN host galaxies to their normal counterparts meaningless \citep{2008ApJ...683..644S}. At high redshifts ($z \gtrsim 0.7$), it is difficult to separate host galaxy from central source even with the superb angular resolution of the \textit{Hubble Space Telescope}. At very low redshifts however, ground-based optical imaging data become comparable, even superior to, the best \textit{Hubble} images of high redshift galaxies. 

We use GALFIT v. 2.0.3 \citep{2002AJ....124..266P}, modified by \cite{2008ApJ...683..644S} for the optimal removal of central point sources, to reliably and robustly separate the central AGN light from the extended host galaxy. In Figure 1, we show two examples of the results of this process. We remove from our sample any sources where the point source is brighter than the host galaxy and sources where the host half-light radius is less than two pixels in order to robustly measure the host galaxy colors down to $\sim$0.1 mag. This leaves us with 16 sources from the \textit{Swift} BAT/SDSS sample. This criterion introduces a slight bias against the most luminous AGN in the BAT sample ($L_{\rm X} \gtrsim 10^{44}  ~\rm erg~s^{-1}$).

\begin{figure*}
\begin{center}

\includegraphics[angle=90, width=\textwidth]{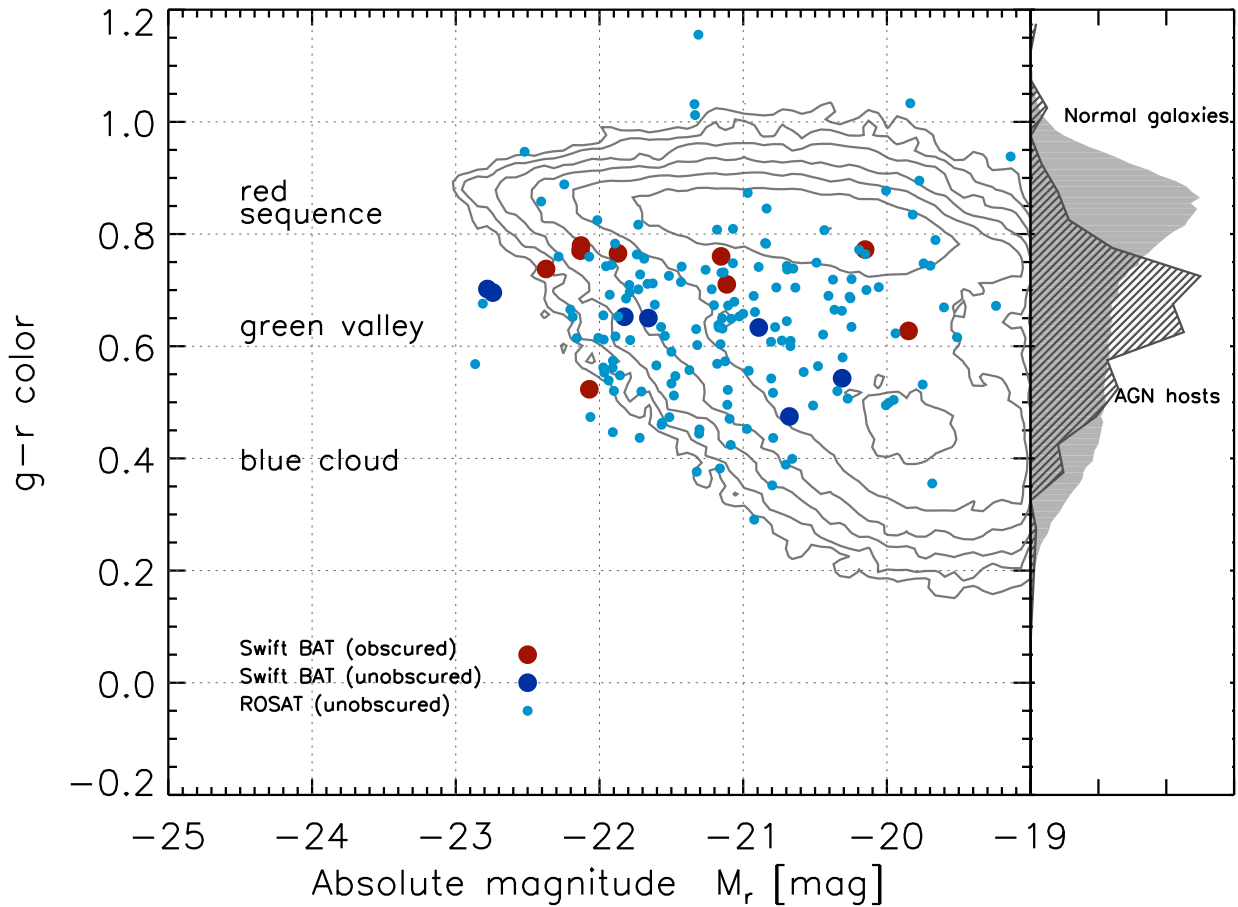}
\caption{The color-magnitude diagram for the host galaxies of X-ray-selected AGN (filled points) and a matched sample of normal galaxies (contours, doubling at each level). The $g-r$ color is a good tracer of the age of the stellar population dominating the optical light. The galaxy comparison sample from SDSS clearly shows the red sequence and blue cloud. After removing the central point source, the X-ray selected AGN appear in the green valley in between. The large points are hard X-ray selected AGN detected by \textit{Swift} BAT, colored according to obscuration ($\log{N_{\rm H}} \leq 22$ -blue; $\log{N_{\rm H}} > 22$ - red). The small points are the unobscured AGN detected by ROSAT. On the right-hand, we plot $g-r$ color histograms of both the galaxy comparison sample and the AGN sample.}

\label{fig:cmr}

\end{center}
\end{figure*}

\section{Results}
\subsection{The Optical Colors of Normal Galaxies and Hard X-ray selected AGN Hosts}
In order to compare AGN host galaxies to normal galaxies, we construct a volume-limited sample of galaxies from SDSS in the same redshift range with $M_{r} < -19$ mag to select galaxies sampling the entire luminosity range of the AGN host galaxies. In Figure 2, we show the color-magnitude diagram for the normal galaxy population, featuring the color bimodality of the red sequence, blue cloud and the green valley in between. On top of this, we plot the host galaxy population of hard X-ray selected AGN (large filled circles). Obscured (red) and unobscured AGN (blue) clearly reside in the green valley. No AGN of similar hard X-ray luminosity, \textit{regardless of obscuration} lies in the region typical of star-forming galaxies in the blue cloud.

We use a Kolmogorov-Smirnov (KS) test to assess the significance of the difference between the host galaxy colors of the comparison sample and the AGN host galaxies and find that they are different at very high significance ($\gg 3\sigma$). In order to strengthen this statement, we compare  the color distribution in luminosity bins. In all four bins of 1 magnitude each between $M_{r}=-23$ to $M_{r}=-19$, the $g-r$ color distribution of AGN host galaxies and of their normal counterparts are found to be significantly different at far more than the $3\sigma$ level.

\subsection{Locating the Obscuring Material}
The unobscured AGN show the same green host galaxy colors as the obscured AGN, meaning the obscuration must be located in the nucleus, not distributed across the galaxy. A KS test of the $g-r$ color distribution of the obscured and unobscured BAT sources yields that they are consistent with being drawn from the same parent distribution at the $2\sigma$ level. The obscuration seen by the AGN must therefore be circumnuclear, as in the Unified Model of AGN, such that the galaxy colors are not strongly affected by the obscuration. \cite{2006ApJ...652L..79T} show that there is evidence that this may not be the case at higher redshifts, towards the peak of AGN and galaxy merger activity, implying evolution of the obscuration.

\subsection{The Optical Colors of Soft X-ray Selected AGN}
Clearly, unobscured AGN appear to have the same green host galaxy colors as obscured AGN -- but the sample is small. To improve statistics, we obtain a larger sample of unobscured AGN using data from the \textit{R\"{o}ntgen Satellite} (ROSAT; \citealt{1999A&A...349..389V, 2007AJ....133..313A}), which is sensitive in the 0.1 to 2.4 keV energy range, and thus to mostly unobscured AGN. Restricting ourselves to the same SDSS area and redshift range, we obtain a sample of 161 unobscured AGN with soft X-ray luminosities of $L_{0.1-2.4 ~\rm keV}$ of $10^{42}-10^{44.3} ~\rm erg~s^{-1}$ and a median of $10^{42.9} ~\rm erg~s^{-1}$. We limit our selection to a minimum of $10^{42} ~\rm erg~s^{-1}$ to avoid contamination with X-rays from star formation and X-ray binaries. The energy ranges observed by \textit{Swift} and ROSAT are different; however, assuming an unobscured AGN spectrum with $\Gamma=1.7$, the median luminosities of our ROSAT and \textit{Swift} BAT samples are comparable within a factor of 3. A KS test affirms that they are consistent with being drawn from the same parent distribution, but note that the  \textit{Swift} BAT sample is small.

\begin{figure*}
\begin{center}

\includegraphics[angle=90, width=0.75\textwidth]{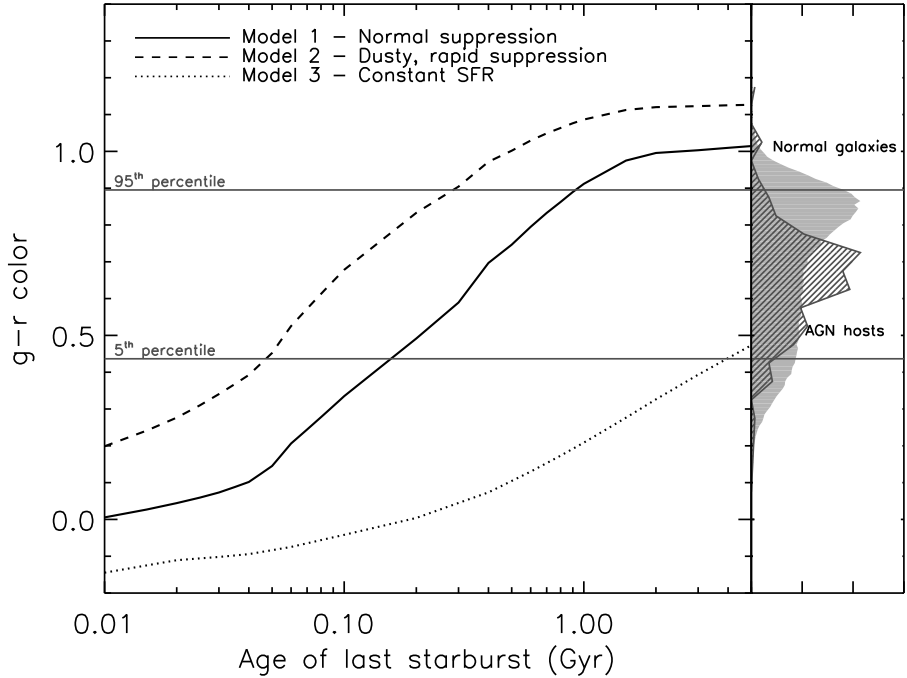}
\caption{Time scales for  `green' optical colors of AGN host galaxies. We plot three star formation scenarios probing the limits of possible color evolution based on stellar evolution models \citep{2005MNRAS.362..799M}. Model 1 is a moderate 10\% mass-fraction starburst on top of an old 8 Gyr stellar population. This burst is somewhat extinguished with E(B-V) = 0.1 and declines with an e-folding timescale of 100 Myr. Model 2 represents an extreme scenario of a small 5\% mass fraction declining effectively instantaneously, with $\tau = $ 10 Myr, extinguished with the maximum allowable E(B-V)= 0.2. Model 3 has a constant star formation history spanning 10 Gyr with E(B-V) of 0.05; such galaxies never leave the blue cloud in most cases. That 95\% of AGN host galaxies lie above $g-r = 0.45$ means that Model 3 is ruled out definitely. Models 1 and 2 are consistent with our results, but imply implausibly long AGN lifetimes unless there is a significant time delay between the suppression of star formation and the onset of detectable AGN activity.}

\label{fig:color}

\end{center}
\end{figure*}

We again subtract the optical central point sources and add the remaining 161 sources with robust colors to Figure 2 (small blue points). Like the unobscured AGN detected by \textit{Swift} BAT, the ROSAT-detected AGN have green host galaxy optical colors. The larger numbers allow us to probe the bluest and reddest colors of AGN host galaxies. and in particular, to see how many objects lie in the blue cloud. We see no AGN in the peak of the blue star-forming galaxy population (lower right of Figure 2). Due to the range of obscuration represented in our sample, this statement must hold for all local moderate-luminosity AGN.

 Could we be systematically missing accreting black holes at the lowest galaxy mass covered by our sample at $M_{r} \sim -19$? At the bluest colors, this corresponds to $\sim 10^{10} ~\rm M_{\odot}$. We convert this to a black hole mass \citep{2004ApJ...604L..89H} and assume that it radiates at efficiencies as low as a few percent of the Eddington limit. Such an AGN would emit a soft X-ray luminosity of $\sim 10^{42} ~ \rm erg~s^{-1}$ and thus should have been detected by ROSAT. The lack of blue AGN hosts is therefore significant and we conclude that the majority of AGN host galaxies in the local universe are in the green valley.

\subsection{A High Merger Fraction for X-ray Selected AGN Host Galaxies}
After subtracting the galaxy model with GALFIT, the residual image contains both the central AGN point source, and any remaining morphological disturbances, such as tidal tails, spiral arms and shells (e.g. Figure 1 top). We perform the same galaxy model subtraction to a sample of normal galaxies matched to the AGN sample in both $g-r$ color and luminosity, and compare the incidence of features indicating a recent or in-progress merger in both. We find a statistically highly significant excess of such features in the AGN host galaxy sample over the control sample (43\% for BAT sources, 21\% for control). We further measure the asymmetry \citep{2003ApJS..147....1C} in both the $g$- and $r$-band images and find that the AGN host galaxies are significantly more asymmetric than those in the control sample. This observation compares favorably to models in which AGN phases are driven by mergers (e.g. \citealt{2006ApJS..163....1H}).

\begin{figure*}
\begin{center}

\includegraphics[angle=90, width=0.75\textwidth]{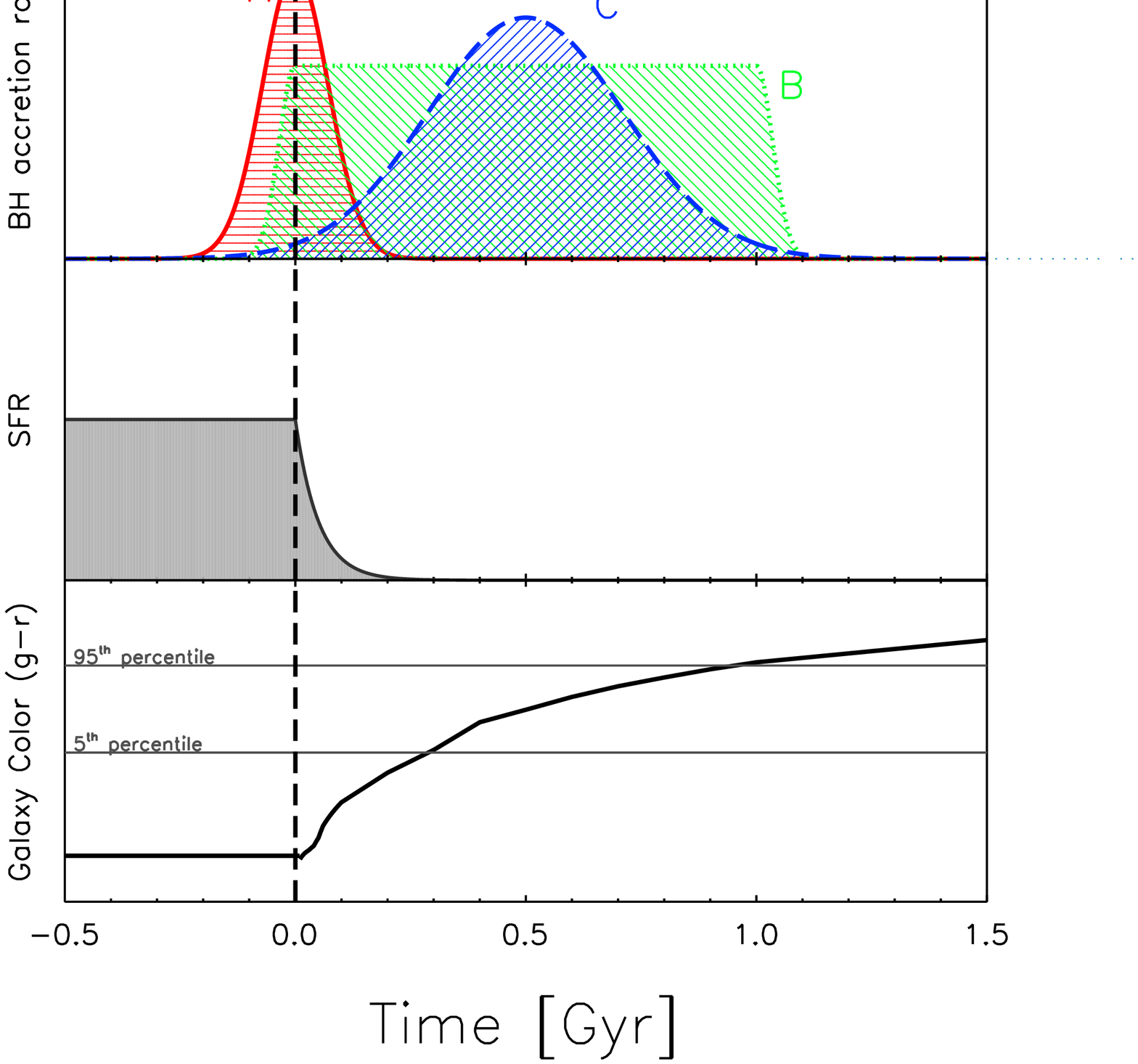}
\caption{In this schematic  diagram, we show the black hole accretion rate, star formation rate and optical color as a function of time, for the case of rapid ($\tau = 100$ Myr) suppression of star formation. In the top panel, we show three Scenarios for the black hole accretion rate. Scenario A \textit{(red, solid)} represents a short-lived AGN/quasar phase that suppresses star formation instantly. This kind of scenario would predict a large number of AGN in the blue cloud, and so is ruled out by our observations. In Scenario B \textit{(green, dotted)}, the AGN quickly switches on, suppresses star formation and remains at high luminosity for $\sim$ 1 Gyr to account for the high fraction of AGN in the green valley compared to the blue cloud. In Scenario C \textit{(blue, dashed)}, the moderate luminosity AGN (like those we detect in the X-rays) does suppress star formation, but peaks in luminosity several 100 Myr after the suppression, naturally giving rise to the observed distribution of AGN host galaxy colors. Such large time delays have been seen in some cosmological simulations (e.g., \citealt{2007ApJ...665..187L}), although these are for high-luminosity quasars at very high redshift.}

\label{fig:diagram}

\end{center}
\end{figure*}

\section{Stellar Population Analysis}
We can use stellar population synthesis as a clock to interpret the green valley optical colors. In Figure 3, we plot $g-r$ colors of three representative star formation histories as a function of time, based on the stellar population models of \cite{2005MNRAS.362..799M}. Models 1 and 2 trace the evolution of realistic and extreme star formation shutdown scenarios respectively, and Model 3 follows a constant star formation rate. Varying the model parameters such as mass-fraction, dust and timescale $\tau$ fills in the parameter space between Models 2 and 3, clustering around the `normal' Model 1.

A particular concern is whether we can account for the green color of AGN host galaxies by invoking substantial ongoing, but obscured, star formation. Any extinction seen by the host galaxy would also be seen by the central engine. The unobscured AGN detected with ROSAT all have column densities of $N_{\rm H} < 10^{22} ~\rm cm^{-2}$. Converting this column density to a dust extinction using the observed ratio of \cite{1978ApJ...224..132B}, $A_{\rm V} = 0.58 (\frac{N_{\rm H}}{10^{21}}$), yields a maximum value for the optical extinction E(B-V) of $\sim 0.2$ for a typical $R_{\rm V} = 3.1$. Substantial ongoing star formation requires sufficient amounts of dust, but such high levels of dust obscuration cannot be present in our AGN host galaxies.

With Model 2, we put a lower bound on the transition time from the blue cloud to the green valley. Longer suppression timescales directly result in longer time delays to the green valley. In Model 1, the $\tau$ of 100 Myr is comparable to the dynamical timescale of massive galaxies and thus sets a natural timescale that compares favorably with observed shutdown timescales inferred in post-starburst galaxies (e.g., \citealt{2007MNRAS.382..960K, 2008arXiv0810.5122W}) and for molecular gas reservoir destruction \citep{2009ApJ...690.1672S}.  The extreme Model 2 assumes virtually instantaneous suppresion ($\tau = 10$ Myr).

The mass fraction of young stars is also a factor in estimating the transition time. The mass-fraction of young stars for the extreme Model 2 of 5\% is chosen to reflect a minor starburst; smaller starburst at the sub-1\% level are widespread, but fail to move red galaxies off the optical red sequence \citep{2005ApJ...619L.111Y, 2007ApJS..173..512S, 2007ApJS..173..619K}. While there are still parameter choice that might give green optical colors at very young ages, such fine tuning seems implausible for our sample of almost 200 objects. The mass-fractions chosen for the extreme (5\%) and realistic (10\%) Models are commensurate to those in green valley early-type galaxies transitioning from the blue cloud to the red sequence \citep{2007MNRAS.382.1415S}. 

For the extreme Model 2, the $g-r$ colors of the AGN host galaxies restrict the minimum time elapsed since the suppression event to $\sim$60 Myr, as only the bluest 5\% of the population exhibit colors that correspond to younger ages. For the more realistic Model 1, the bluest 5\% imply typical ages of a few hundred Myr.

\section{Discussion}
The fact that we detect very few AGN in the peak of the blue cloud seems to rule out scenarios where a short-lived, luminous AGN/quasar phase episode suppresses star formation on short timescales (i.e. Scenario A (red) in Figure 4). 

Two scenarios instead can account for the lack of young, blue AGN host galaxies. The first is that the AGN suppress star formation in their hosts, but the typical AGN lifetime is long compared to the suppression timescale, so that most AGN have evolved away from the blue cloud  (Scenario B (green) in Figure \ref{fig:diagram}). Roughly 5\% of our AGN still have ages less than $\sim$60 Myr, in which case Scenario B implies a lifetime of $\gtrsim$1.2 Gyr. However, the estimates of \cite{2004MNRAS.351..169M} reach $\gtrsim$1 Gyr only for the most massive black holes. The current consensus in the literature (e.g., \citealt{2001ApJ...547...12M, 2004cbhg.symp..169M,2004MNRAS.351..169M}) give typical lifetimes of only a few $10^{8}$ years, which is smaller than the minimum lifetime implied by the dearth of blue host galaxies. Such lifetime estimates are generally given for accretion phases with high Eddington ratios, with the esimate of a few $10^{8}$ years of  \cite{2004MNRAS.351..169M} considering efficiencies down to 0.1 $L_{\rm Edd}$. The median \textit{Swift} BAT source in our sample has efficiencies of $L \sim 0.1 L_{\rm Edd}$ with a large scatter if we compute  $L/L_{\rm Edd}$, for black hole masses estimated from the $M_{\bullet} - \sigma$ relation \citep{2000ApJ...539L...9F, 2000ApJ...539L..13G} and apply a bolometric correction from $L_{\rm X}$ to $L_{\rm bol}$ of 20 to 100.

The other interpretation (Scenario C (blue) in Figure 4), is that star formation is suppressed immediately, but the AGN luminosity is detectable only with a delay of about 100 Myr. If so, low-luminosity AGN during the rising phase, below $\sim10^{42} ~\rm erg~s^{-1}$ , must be sufficient to shut down star formation, as suggested by the destruction of cold molecular gas reservoirs by low-luminosity AGN \citep{2007MNRAS.382.1415S, 2009ApJ...690.1672S}. Then the optical colors of the green valley AGN are a reflection of the time needed for substantial amounts of material to lose enough angular momentum to reach the black hole and thus increase the AGN luminosity. This time gap of $\sim100$ Myr cannot be accounted for by a heavily obscured AGN phase, as \textit{Swift} BAT would have detected them. In this Scenario C, the low-luminosity phase and not the X-ray bright phase is the cause of the shutdown of star formation. Perhaps the low-luminosity phase, while radiatively inefficient, is channeling substantial amounts of kinetic energy into the ISM, similar to the `radio' or `maintenance 'mode invoked in current semi-analytic models \citep{2006MNRAS.370..645B, 2006MNRAS.365...11C, 2008MNRAS.tmp.1241S}. This Scenario also leaves the option that another process causes both the shutdown of star formation and the AGN phase with a time delay, but such a scenario raises more problems than it solves.

Our observational result challenges recent results from simulations of AGN feedback, at least at low redshifts \citep{2005MNRAS.361..776S, 2005Natur.433..604D, 2006ApJS..163....1H, 2008arXiv0802.0210J}, which suggest that the epoch of maximum black hole growth is coeval with the peak of star formation. The coeval scenario (i.e. Scenario A) where the AGN luminosity peaks during the suppression phase, would predict a large population of high-luminosity AGN in the blue cloud, where we observe very few. We note however that some of these simulations pertain to the high redshift universe, and that the physical processes at low redshift may be different.

\acknowledgements We thank Joseph Silk, Yuexing Li and Sukyoung Yi for comments and discussions. SV acknowledges support from a graduate research scholarship awarded by the Natural Science and Engineering Research Council of Canada (NSERC) and from NASA/INTEGRAL grant NNG05GM79G. BS and CMU received support from NSF grant AST0407295. Support for ET was provided by the National Aeronautics and Space Administration through Chandra Postdoctoral Fellowship Award Number PF8-90055 issued by the Chandra X-ray Observatory Center, which is operated by the Smithsonian Astrophysical  Observatory for  and on behalf of the National Aeronautics Space Administration under contract NAS8-03060. SK acknowledges a research fellowship from the Royal Commission for the Exhibition of 1851. This publication is based on data from the Sloan Digital Sky Survey, the \textit{Swift} Gamma Ray Burst Telescope and the \textit{R\"{o}ntgen Satellite} (ROSAT). \\

{\it Facilities:} \facility{\textit{Swift} (BAT), ROSAT(), Sloan()}

\bibliographystyle{astroads}

 \newpage

\clearpage

\clearpage

\clearpage

\end{document}